\documentstyle[12pt]{article}
\begin{document}
\begin{center}{\large{\bf The motion of a charged particle in Kaluza-Klein manifolds}}
\end{center}
\vspace*{1.5cm}
\begin{center}
A. C. V. V. de Siqueira
$^{*}$ \\
Departamento de Educa\c{c}\~ao\\
Universidade Federal Rural de Pernambuco \\
52.171-900, Recife, PE, Brazil.\\
\end{center}
\vspace*{1.5cm}
\begin{center}{\bf Abstract}

In this paper we use Jacobi fields to describe the motion of a
charged particle in the classical gravitational, electromagnetic,
and Yang-Mills fields.
\end{center}

 \vspace{3cm}

${}^*$ E-mail: acvvs@ded.ufrpe.br
\newline

\newpage

\section{Introduction}
$         $

The Jacobi fields are very important to the Riemannian
Geometry\cite{1} and in singularity theorems\cite{2}. These fields
can also be used to study the motion of a particle in a Riemannian
spacetime as an alternative to the usual geodesic approach. The
general relativity theory assures us that every coordinate frame
is physically equivalent from a classical point of view. Thus we
can claim that any falling particle in a gravitational field will
be accelerated relatively to any stationary coordinate frame
defined by a Killing vector. However, it is possible that a
specific spacetime does not have Killing vector, so in this case
we would not have at our disposal a frame which enables us to
measure this acceleration. In such a situation, the best we can do
is to take two different particles and measure their relative
acceleration. This is a relativistic three-body problem. The
metric deformation by test particles\cite{3} is necessary in a
more realistic approach to the Kaluza-Klein manifolds, but it is a
difficult problem in geometry so that we will not consider it. In
this case a relative acceleration will be given by the equation
which governs the Jacobi fields.

 This paper is organized as follows. In Sec.$2$ we present some facts
about the Jacobi fields. In Sec.$3$  we build a Kaluza-Klein
geometry and the associated  Jacobi equation. In Sec.$4$ we
summarize the main results of this work.

\renewcommand{\theequation}{\thesection.\arabic{equation}}
\section{\bf Jacobi Fields}
\setcounter{equation}{0}
$         $

In this section we briefly review the Jacobi fields and their
respective differential equation for a Riemann manifold. Let us
consider a differentiable manifold, $\cal{M}$, and two structures
defined on $\cal{M}$, namely affine connection, $\nabla$ , and
Riemann tensor, $K$ , related by the equation\cite{2}

\begin{equation}
 K(X,Y)Z=\nabla _{X}\nabla _{Y}Z-\nabla _Y\nabla
_{X}Z-\nabla _{[X,Y]}Z,
\end{equation}
where $X,Y$ and $Z$ are vector fields in the tangent space. The
torsion tensor can be defined by
\begin{equation}
 T(X,Y)=\nabla _{X}Y-\nabla _{Y}X-[X,Y],
\end{equation}
where $\nabla _{X}Y$ is the covariant derivative of the field $Y$
along the $X$ direction. We  define the Lie derivative by

\begin{equation}
 {\cal{L}}_{X}Y=[X,Y].
\end{equation}
For a torsion-free connection we can write

\begin{equation}
 {\cal{L}}_{X}Y=\nabla _{X}Y-\nabla _{Y}X.
\end{equation}
Let us consider the tangent field V, the Jacobi fields Z and the
condition
\begin{equation}
 {\cal{L}}_{V}Z=0.
\end{equation}
From (2.5), we can easily see that the equations which govern the
Jacobi field can be written in the form

\begin{equation}
 \nabla _V\nabla _VZ+K(Z,V)V-\nabla _Z\nabla _VV=0.
\end{equation}
We assume that
\begin{equation}
 \nabla _{V}V=0.
\end{equation}
In this case the Jacobi equations are reduced to a geodesic
deviation, assuming a simpler form, and the Fermi derivative
$\frac{D_FZ}{\partial s}$ coincides with the usual covariant
derivative

$$
\frac{D_FZ}{\partial s}=\frac{DZ}{\partial s}=\nabla _VZ,
$$
and

\begin{equation}
 \nabla _V\nabla _VZ+K(Z,V)V=0.
\end{equation}

\renewcommand{\theequation}{\thesection.\arabic{equation}}
\section{\bf Jacobi equation in a Kaluza-Klein manifold}

\setcounter{equation}{0}
$         $

We intend to make use of a vielbein basis, but it is necessary,
beforehand, to present a local Kaluza-Klein metric G for the
fields\cite{4,5,6}. The metric tensor G has signature
$(-,+,\ldots,+).$ For Kaluza-Klein manifold the indices
$\Lambda,\Pi$ $\in$ $(0,1,2,3,4,5,6,\ldots,n)$ and for ordinary
space-time the indices \linebreak $\alpha, \beta$ $ \in$ $
(0,1,2,3).$ For internal space we have indices $x^{4}=y$ for the
electromagnetic fields and $i,j \in (5,6,\ldots,n)$ for the
Yang-Mills fields. We can write a local basis coordinate system as
$x^{\Lambda}=(x^{\alpha},y,z^{i}).$ In this basis the metric G is
given by the ansatz
\begin{eqnarray}
 && G_{\Lambda \Pi}=g_{\alpha \beta}+\eta_{(4)
(4)}A_{\alpha}A_{\beta} \\ \nonumber && + h_{i
j}\xi^{i}_{m}\xi^{j}_{n}A^{m}_{\alpha}A^{n}_{\beta},
\end{eqnarray}
\begin{equation}
 G_{\alpha 4}=\eta_{(4)(4)}A_{\alpha},
\end{equation}
\begin{equation}
 G_{i 4}=0,
\end{equation}
\begin{equation}
G_{\alpha i}=h_{i j}\xi^{i}_{m}A^{m}_{\alpha},
\end{equation}
where the form invariance of $h_{i j}(z)$ was assumed and this
implies that Killing equation is obeyed, or equivalently
\begin{equation}
L_{\xi}h=0.
\end{equation}
Next, let us consider the connection between the vielbein and the
local metric tensor
\begin{equation}
G_{\Lambda\Pi}=E_{\Lambda}^{(\mathbf{A})}E_{\Pi}^{(\mathbf{B})}\eta_{(\mathbf{A})(\mathbf{B})},
\end{equation}
where $ \eta_{(\mathbf{A})(\mathbf{B})}$ and $
E_{\Lambda}^{(\mathbf{A})}$ are Lorentzian metric and vielbein
components respectively. The flat indices
$(\mathbf{A}),(\mathbf{B}),\ldots,(\mathbf{M}),(\mathbf{N})$ $\in$
$(0,1,2,3,4,5,6,\ldots,n).$ Specifically we have $(\mathbf{A})$
$=(A,{(4)},a),$ with  $A \in$ $ (0,1,2,3),$ $a \in$
$(5,6,\ldots,n),$ so that the Lorentzian metric is composed as
follows
\begin{equation}
 \eta_{(\mathbf{A})(\mathbf{B})}=(\eta_{A B},\eta_{(4)(4)},\eta_{a b}),
\end{equation}
where $\eta_{A B}$ are the flat metric of the ordinary space-time
and $ \eta_{(4)(4)},$ $\eta_{a b}$ of the internal space, are
associated with the electromagnetic and Yang-Mills fields
respectively. In the vielbein basis we use the following
Riemannian curvature
\begin{eqnarray}
&& K_{(\mathbf{A})(\mathbf{B})(\mathbf{C})(\mathbf{D})}=-\gamma_{(\mathbf{A})(\mathbf{B})(\mathbf{C}),(\mathbf{D})}+\gamma_{(\mathbf{A})(\mathbf{B})(\mathbf{D}),(\mathbf{C})}\\
\nonumber
&& +\eta^{(\mathbf{M})(\mathbf{N})}[\gamma_{(\mathbf{B})(\mathbf{A})(\mathbf{M})}(\gamma_{(\mathbf{C})(\mathbf{N})(\mathbf{D})}-\gamma_{(\mathbf{D})(\mathbf{N})(\mathbf{C})}) \\
\nonumber &&
+\gamma_{(\mathbf{M})(\mathbf{A})(\mathbf{C})}\gamma_{(\mathbf{B})(\mathbf{N})(\mathbf{D})}-\gamma_{(\mathbf{M})(\mathbf{A})(\mathbf{D})}\gamma_{(\mathbf{B})(\mathbf{N})(\mathbf{C})}],
\end{eqnarray}
where the
$\gamma_{(\mathbf{A})(\mathbf{B})(\mathbf{C})(\mathbf{D})}$ are
the Ricci rotation coeficients.
 The only non-vanishing coeficients are
\begin{equation}
 \gamma_{a B C}=1/2\xi_{m a}F^{m}_{A B} ,
\end{equation}
\begin{equation}
 \gamma_{A B c} = -1/2\xi_{m c}F^{m}_{A B} ,
\end{equation}
\begin{equation}
 \gamma_{(4) A B}= 1/2\eta_{(4)(4)}F_{A B} ,
\end{equation}
\begin{equation}
 \gamma_{A B (4)}= -1/2\eta_{(4)(4)}F_{A B} ,
\end{equation}
\begin{equation}
 \gamma_{A B C}=E_{A}^{\Lambda}E_{\Lambda B;\Pi}E_{C}^{\Pi},
\end{equation}
with
\begin{equation}
 F_{A B}= A_{B,A}-A_{A,B},
\end{equation}
and
\begin{equation}
 F^{m}_{A B}= A^{m}_{B,A}-
 A^{m}_{A,B}+f^{m}_{ln}A^{l}_{A}A^{n}_{B},
\end{equation}
where $A_{A}=A_{A}(x),$  $ A^{m}_{A}= A^{m}_{A}(x),$
$\xi_{ma}=\xi_{ma}(z),$ and $\xi_{ma}=\eta_{mb}\xi^{b}_{a}.$ Using
the above results in (3.8) we obtain the following Riemannian
components
\begin{eqnarray}
&& K_{ABCD}= R_{ABCD}+1/4\eta_{(4)(4)}[ -2F_{AB}F_{CD}
-F_{AC}F_{BD}\\ \nonumber && + F_{AD}F_{BC}]
 +1/4\eta^{ab}\xi_{ma}\xi_{lb}[-2F^{m}_{AB}F^{l}_{CD}
  \\ \nonumber && -F^{m}_{AC}F^{l}_{BD} + F^{m}_{AD}F^{l}_{BC}],
\end{eqnarray}

\begin{eqnarray}
 &&K_{abCD}=\gamma_{abd}\xi^{d}_{m}F^{m}_{CD}\\ \nonumber
 && - 1/4 [\xi_{l a}\xi_{m b} - \xi_{l b}\xi_{m a} ]\eta^{M N}F^{l}_{NC}F^{m}_{MD},
\end{eqnarray}

\begin{equation}
K_{(4)B(4)C}=1/4\eta^{2}_{(4) (4)}\eta^{M N}F_{MB}F_{N C},
\end{equation}

\begin{eqnarray}
&& K_{(4)BCD}=1/2 \eta_{(4) (4)}(F_{BD,C} - F_{BC,D})\\ \nonumber
&& + \eta^{M N } [(\gamma_{DNC} - \gamma_{CND} ) F_{BM} -
\gamma_{BND}F_{MC} + \gamma_{BNC}F_{MD}],
\end{eqnarray}

\begin{eqnarray}
&& K_{aBCD}=1/2\xi_{ma}\{F^{m}_{BC,D}- F^{m}_{BD,C} + \eta^{MN}
[(\gamma_{DNC} \\ \nonumber && - \gamma_{CND})F^{m}_{BM} +
\gamma_{BNC}F^{m}_{MD} - \gamma_{BND}F^{m}_{MC}]\} ,
\end{eqnarray}
where $R_{ABCD}(x)$ are the Riemannian components in the ordinary
spacetime. It is important to note that the following properties
can be verified
\begin{equation}
K_{(\mathbf{A})(\mathbf{B})(\mathbf{C})(\mathbf{D})}=K_{(\mathbf{C})(\mathbf{D})(\mathbf{A})(\mathbf{B})},
\end{equation}
and
\begin{equation}
K_{(\mathbf{A})(\mathbf{B})(\mathbf{C})(\mathbf{D})} +
K_{(\mathbf{A})(\mathbf{D})(\mathbf{B})(\mathbf{C})}+K_{(\mathbf{A})(\mathbf{C})(\mathbf{D})(\mathbf{B})}=0,
\end{equation}
where for some set of indices we have used (3.5).

We now consider a massive test particle. Since the particle has a
non-vanishing rest mass, it is convenient to define the tangent
vector $V$ as a timelike one, so that $g(V,V)=-1.$ Let us build a
Fermi-Walker transport. In the Fermi-Walker transported particle
frame the equation (2.8) is given by
\begin{equation}
\frac{d^2Z_{(\mathbf{A})}}{d\tau
^2}+K_{0(\mathbf{A})0(\mathbf{C})}Z_{(\mathbf{C})}=0.
\end{equation}
More specifically:
\begin{equation}
\frac{d^2Z_{\tilde{A}}}{d\tau^{2}} + K_{0 \tilde{A}0
\tilde{C}}Z_{\tilde{C}} + K_{0 \tilde{A} 0 (4)}Z_{(4)} + K_{0
\tilde{A} 0 a}Z_{a}=0,
\end{equation}
\begin{equation}
\frac{d^2Z_{(4)}}{d\tau ^2}+K_{0(4)0(4)}Z_{(4)}=0,
\end{equation}
 and
\begin{equation}
\frac{d^2Z_{a}}{d\tau^{2}} + K_{0a0b}Z_b + K_{0 a 0
\tilde{C}}Z_{\tilde{C}}=0,
\end{equation}
where $\tau $ is, in general, an affine parameter, which in our
case is the proper time of the particle and $Z_{(\mathbf{A})}$ are
the vielbein components of the space-like vector $Z$, with
$g(Z,V)=0$ and $\tilde{A},\tilde{C}=(1,2,3).$
\newpage
\section{Concluding Remarks}
  $              $
The equations (3.24), (3.25), and (3.26) indicate that the
particle motion is a function of the gravitational,
electromagnetic and Yang-Mills fields. The equation (3.24)
suggests that the particle motion in a ordinary spacetime can
depend on matter constituent details, but this effect is very
small when compared to the others. We do not expect that
(3.24)-(3.26) are a representation of the complete ansatz. The
metric deformation by test particles\cite{3} is necessary,
although it is not enough in a more realistic approach. Actually,
it is an open problem in geometry. We believe that the results
obtained in the present paper might lead to a better comprehension
of these issues in the future. \footnote{ We are grateful to Dr.
Marcelo M. Leite and Dr. Cat\~ao Barbosa, for their useful
discussions.}

\end{document}